\documentclass[10pt, preprint]{emulateapj}
\def\arcs{$''$}

\slugcomment{submitted to ApJ Supplements}

\newcommand{\myemail}{zfan@bao.ac.cn}

\shorttitle{Star clusters in M33} \shortauthors{Z. Fan \& R. de Grijs}

\begin{document}

\title{Star clusters in M33: updated $UBVRI$ photometry, 
  ages, metallicities, and masses}

\author{Zhou Fan\altaffilmark{1} and Richard de Grijs\altaffilmark{2,3}}

\altaffiltext{1}{Key Laboratory of Optical Astronomy, National
  Astronomical Observatories, Chinese Academy of Sciences, 20A Datun
  Road, Chaoyang District, Beijing 100012, China}

\altaffiltext{2}{Kavli Institute for Astronomy and Astrophysics,
  Peking University, Yi He Yuan Lu 5, Hai Dian District, Beijing
  100871, China}

\altaffiltext{3}{Department of Astronomy, Peking University, Yi He
  Yuan Lu 5, Hai Dian District, Beijing 100871, China}

\email{\myemail, grijs@pku.edu.cn}

\begin{abstract}
The photometric characterization of M33 star clusters is far from
complete. In this paper, we present homogeneous $UBVRI$ photometry of
708 star clusters and cluster candidates in M33 based on archival
images from the Local Group Galaxies Survey, which covers 0.8 deg$^2$
along the galaxy's major axis. Our photometry includes 387, 563, 616,
580, and 478 objects in the $UBVRI$ bands, respectively, of which 276,
405, 430, 457, and 363 do not have previously published $UBVRI$
photometry. Our photometry is consistent with previous measurements
(where available) in all filters. We adopted Sloan Digital Sky Survey
$ugriz$ photometry for complementary purposes, as well as Two Micron
All-Sky Survey near-infrared $JHK$ photometry where available. We
fitted the spectral-energy distributions of 671 star clusters and
candidates to derive their ages, metallicities, and masses based on
the updated {\sc parsec} simple stellar populations synthesis
models. The results of our $\chi^2$ minimization routines show that
only 205 of the 671 clusters (31\%) are older than 2 Gyr, which
represents a much smaller fraction of the cluster population than that
in M31 (56\%), suggesting that M33 is dominated by young star clusters
($<1$ Gyr). We investigate the mass distributions of the star
clusters---both open and globular clusters---in M33, M31, the Milky
Way, and the Large Magellanic Cloud. Their mean values are
$\log(M_{\rm cl}/M_{\odot})=4.25$, 5.43, 2.72, and 4.18,
respectively. The fraction of open to globular clusters is highest in
the Milky Way and lowest in M31. Our comparisons of the cluster ages,
masses, and metallicities show that our results are basically in
agreement with previous studies (where objects in common are
available); differences can be traced back to differences in the
models adopted, the fitting methods used, and stochastic sampling
effects.
\end{abstract}


\section{Introduction}
\label{s:intro}

Since star clusters represent an important component of the galaxies
they are associated with, studies of star clusters' stellar
populations and age distributions can provide clues to the formation
and evolution of their host galaxies. In addition, since populous star
clusters are much more luminous than individual stars, they are
usually much easier to observe and study.

At a distance of $847\pm60$ kpc---equivalent to a distance modulus of
$(m-M)_0=24.64\pm0.15$ mag \citep{gal04}---M33 (also known as the
Triangulum Galaxy) is the third largest spiral galaxy in the Local
Group of galaxies. Since the galaxy is seen relatively face-on, under
an inclination of $i=56^{\circ}\pm1^{\circ}$ \citep{zeh}, it is
eminently suitable for studies of its star cluster system. At present,
the most comprehensive and widely used star cluster catalog is that of
\citet{sm07}, which combines data on almost all M33 star clusters
published in the literature, including information on their
photometry, ages, metallicities, and masses. The latest
version\footnote{http://www.mancone.net/m33$\_$catalog/, updated in
  December 2010.} (henceforth SM10) includes 595 star clusters and
candidates. \citet{pl07} found 104 star clusters in {\sl Hubble Space
  Telescope} ({\sl HST})/Wide Field and Planetary Camera-2 (WFPC2)
archival images, including 32 new objects based on new {\sl HST}
observations. Although their observations improved the spatial
coverage of the M33 disk, this catalog is still incomplete for the
entire disk. These authors found two different star cluster
populations on the basis of their sample's color--magnitude diagram
(CMD), including a large number of blue clusters and a smaller number
of red objects. They also suggested that relatively more red clusters
are found in the galaxy's outer regions.

Subsequently, \citet{zkh} published a list of 4780 extended sources,
including 3554 new cluster candidates observed with the MegaCam
instrument on the 3.6 m Canada--France--Hawai'i Telescope
(CFHT). However, $\sim$60\% of these clusters are not considered
genuine owing to possible misidentifications
\citep{san09,san10}. Based on {\sl HST}/Advanced Camera for Surveys
(ACS)--Wide Field Channel (WFC) observations, \citet{san09} presented
photometry of 161 M33 star clusters, of which 115 were newly
identified. Based on their CMDs, they suggested that these clusters'
ages were between 0.01 and 1 Gyr, whereas their masses range from
$5\times10^3 M_\odot$ to $5\times10^4 M_{\odot}$. However these
authors also point out that, since their photometry is generally not
sufficiently deep to detect the main-sequence turnoff (MSTO), very few
of their sample clusters are older than 1 Gyr. Using MegaCam on the
CFHT, \citet{san10} identified 2990 extended sources in M33, 599 of
which were new cluster candidates and 204 were previously known
clusters. Based on CMD analysis, these authors suggested that the
majority of the clusters have young to intermediate ages, although
their sample also includes some old objects. They suggested that a
possible M31--M33 interaction some 3.4 Gyr ago may have triggered an
epoch of star (cluster) formation in M33.

Comparison of observational spectral-energy distributions (SEDs) with
theoretical stellar population synthesis models by application of
$\chi^2$ minimization is a widely used technique to estimate ages,
metallicities, reddening values, and masses of extragalactic star
clusters. This technique has been applied to the cluster systems in,
e.g., M31 \citep{jiang03,fan06,fan10,ma07,ma09,wang10}, M33
\citep{ma01,ma02a,ma02b,ma02c,ma04a,ma04b}, the Large Magellanic Cloud
\citep[LMC; e.g.,][]{da06,pop12,grijs13}, M82 \citep{deg03a,lim13},
NGC 3310 and 6745 \citep{deg03b}, as well as for stellar population
synthesis model comparisons \citep{deg05,fan12}.

In this paper, we first obtain photometry for all M33 star clusters in
our sample (see Section \ref{s:samp} for definition) based on archival
images from the Local Group Galaxies Survey
\citep[LGGS;][]{massey}. Using photometry in the $UBVRI, ugriz$ (Sloan
Digital Sky Survey; SDSS) bands and Two Micron All-Sky Survey (2MASS)
$JHK$ magnitudes
\citep{2mass}\footnote{http://www.ipac.caltech.edu/2mass/} when
available, the ages and masses of the star clusters in our sample are
estimated by comparison of their observed SEDs with updated {\sc
  parsec} (version 1.1) isochrones \citep{bre12}. This paper is
organized as follows. Section \ref{s:data} describes the sample
selection and $UBVRI$ photometry. In Section \ref{s:age} we describe
the simple stellar population (SSP) models used, as well as our method
to estimate the cluster ages and metallicities. In Section
\ref{s:mass} we present the clusters' mass estimates, and we summarize
and conclude the paper in Section \ref{s:sum}.

\section{Data}
\label{s:data}

\subsection{Sample}
\label{s:samp}

Our sample star clusters are mainly selected from \citet{san10}, whose
database is based on observations with the CFHT/MegaCam camera. Their
catalog covering the M33 area contains 2990 objects, including
background galaxies, confirmed star clusters, and cluster candidates,
as well as unknown objects. The catalog provides the positions and
$ugriz$ photometry of all objects. Since our focus is on the star
clusters, galaxies and unknown objects were eliminated from the
catalog, and we subsequently performed photometry for the 803 star
clusters and cluster candidates in their catalog.

We used archival $UBVRI$ images from the LGGS, which covers a region
of 0.8 deg$^2$ along the galaxy's major axis. The images we used
consisted of three separate but overlapping fields with a scale from
0.261$''$ pixel$^{-1}$ at the center to 0.258$''$ pixel$^{-1}$ in the
corners of each image. The field of view of each mosaic image is
$36\times36$ arcmin$^2$. The observations were taken with the Kitt
Peak National Observatory 4 m telescope between August 2000 and
September 2002. The median seeing of the LGGS images is $\sim$
1\arcs. Although \citet{ma12} inspected the images and obtained
$UBVRI$ photometry for all star clusters and unknown objects in
\citet{sm07} based on archival LGGS images, there are still hundreds
of star clusters from \citet{san10} in this field which do not have
published $UBVRI$ photometry. Therefore, here we only perform
photometry of the clusters in the LGGS images following the
identifications of \citet{san10}. We employed the latest version of
{\sc
  SExtractor}\footnote{http://www.astromatic.net/software/sextractor;
  version 2.8.6 was updated on 5 October 2009.} \citep{ba96} to find
the sources in the images and match them to the coordinates of our 803
sample star clusters and candidates. Eventually, we detected 588
clusters and candidates with quality FLAGS = 0, which indicates that
there are no problems associated with these objects (i.e., no
contamination by nearby sources or saturation effects) in the LGGS
images.

To supplement these data, we also include 120 confirmed star clusters
from the updated (2010) version of \citet[][SM10]{sm07}, which were not 
included in \citet{san10}. Thus, the number of clusters in our final 
sample is 708.

\begin{figure}
\centerline{
  \includegraphics[scale=0.5,angle=0]{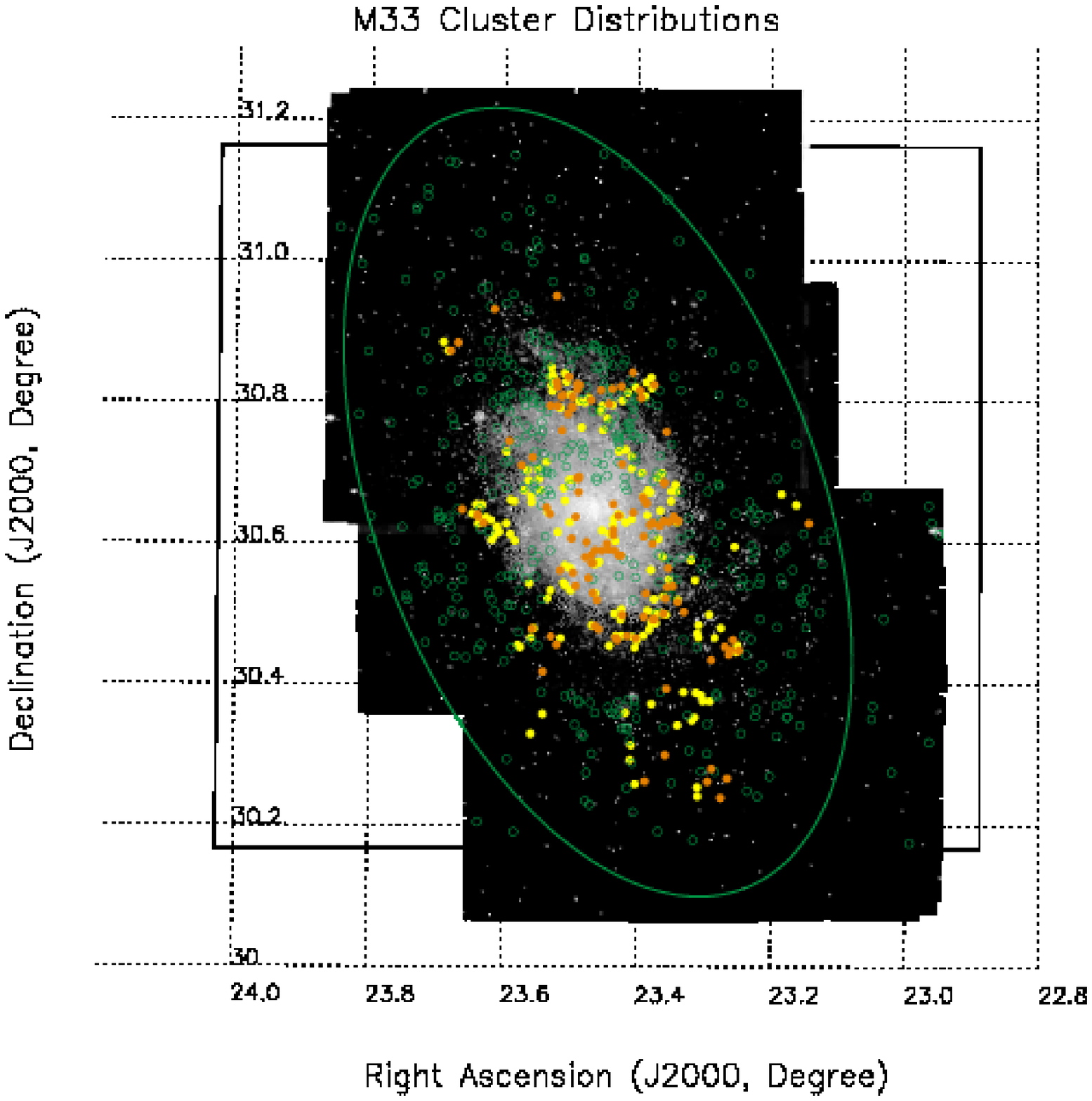}}
\caption[]{Spatial distribution of our star clusters and cluster
  candidates. The yellow solid bullets (confirmed clusters) and
    green open circles (candidate clusters) are from \citet{san10}; we
    determined their cross-identifications in the \citet{massey}
    image. The orange symbols represent the clusters identified by
  SM10 but not included in \citet{san10}. The three data frames are
  the fields of view of \citet{massey}, whereas the large square
  outline covers the observed field of \citet{san10}.}
  \label{fig1}
\end{figure}

Figure \ref{fig1} shows the spatial distribution of all sample
clusters and candidates in the M33 field. The three data frames
represent the field of view of the \citet{massey} data, and the large
square outline covers the observed field of \citet{san10}. The
large green ellipse delineates $D_{25}$, i.e., it corresponds to the
$\mu_B = 25$ mag arcsec$^{-2}$ isophote \citep{boi07}. The yellow
solid bullets and green open circles are, respectively, the
confirmed star clusters and cluster candidates from
\citet{san10}. The combined total number of clusters and cluster
candidates is 588. These latter clusters have been cross-identified
in the \citet{massey} image, which we will focus on here; we do not
consider clusters outside of the boundaries of the \citet{massey}
image. The orange symbols represent the 120 star clusters identified
by SM10 but not included in \citet{san10}. It is clear that most star
clusters and candidates are associated with and projected onto the
galaxy's disk (i.e. inside the galaxy's $D_{25}$).

\subsection{Integrated photometry}
\label{s:phot}

Prior to this work, \citet{massey} compiled point-spread-function
(PSF) photometry for 146,622 stars (point sources) in the M33 fields,
with photometric uncertainties of $<10$\% at $UBVRI \sim 23$
mag. However, there are no relevant discussions of the extended
sources (e.g., star clusters and galaxies) in the published LGGS
papers. Recently, \citet{ma12} derived aperture photometry of 392 star
clusters and unknown objects in the catalog of \citet{sm07} in the
$UBVRI$ bands. However, there are still several hundred M33 star
clusters and cluster candidates identified by \citet{san10} which lack
LGGS photometry in the $UBVRI$ bands.

To obtain additional photometric information for the star clusters, we
carried out photometric measurements of our sample M33 clusters and
candidates. {\sc SExtractor} was applied to the LGGS images in all of
the $UBVRI$ bands to derive supplementary and homogeneous
photometry. The {\sc SExtractor} code provides isophotal magnitudes
corrected for the flux missed by isophotal-magnitude determination,
MAG$\_$ISOCOR. This approach works well for stars but poorly for
elliptical (galaxy) profiles with broader wings. {\sc SExtractor} also
delivers automatic aperture photometry measurements of galaxies based
on the first-moment algorithm of \citet{kr80}, MAG$\_$AUTO. The
MAG$\_$BEST magnitudes can be automatically mapped onto MAG$\_$AUTO if
neighbors cannot bias the photometry by more than 10\%. In all other
cases, MAG$\_$BEST is set to equal MAG$\_$ISOCOR, because the latter
measurements are not significantly affected by nearby sources. Thus,
we adopted the MAG$\_$BEST magnitudes as our final instrumental
magnitudes. As a consequence, we do not need to choose the size of the
aperture used. The instrumental magnitudes were calibrated in the
standard Johnson--Kron--Cousins $UBVRI$ system by comparing the
published magnitudes of stars from \citet{massey}, who calibrated
their photometry using \citet{lan92} standard stars, with our
instrumental magnitudes. Since the magnitudes in \citet{massey} are
given in the Vega system, our photometry is also tied to that
system. The calibration errors range from $\sim0.01$ to $\sim0.03$ mag
in the $UBVRI$ bands, with more than 300 secondary standard stars
available in each field. Finally, we obtained photometry for 708
objects, with 387, 563, 616, 580, and 478 sources in the individual
$UBVRI$ bands, respectively, of which 276, 405, 430, 457, and 363 star
clusters and candidates do not have previously published photometry.

Table 1 of \citet{massey} shows that the seeing conditions under
which the LGGS fields were obtained ranged from $0''.8$ to $1''.2$
in all filters; for most fields the prevailing seeing was around
$1''.0$. In their table 3, these authors compared the differences in
their calibrated photometry between overlapping fields using
well-exposed, isolated stars and found that the median difference
was several millimagnitudes. Our photometry has been calibrated
relative to that of \citet{massey}. We compared the photometric
measurements of those clusters that were located in the regions of
overlap between different frames and and found differences of only a
few $\times 0.01$ mag. While these differences are a little larger
than those reported by \citet{massey}, this is not unexpected, since
star clusters often have more extended and more complicated profiles
than stars. For clusters with more than one photometric measurement
in overlapping fields, we adopted the magnitude associated with the
smallest statistical uncertainty.

Table \ref{photo} lists our new broad-band $UBVRI$ magnitudes and the
corresponding photometric errors. The latter combine the errors
associated with MAG$\_$BEST with those related to the flux
calibration, as
\begin{equation}
  \sigma_i^{2}=\sigma_{{\rm best},i}^{2}+\sigma_{{\rm calib},i}^{2},
\label{eq1}
\end{equation}
where $i$ represents any of the $UBVRI$ bands, whereas $\sigma_{\rm
  best}$ and $\sigma_{\rm calib}$ correspond to the photometric
uncertainties associated with the MAG$\_$BEST magnitudes and flux
calibration, respectively.

Since {\sc SExtractor} applies apertures of different sizes to obtain
MAG$\_$BEST magnitudes, depending on the size of the object of
interest, we applied aperture growth-curve corrections to all
photometric measurements. In fact, although the MAG$\_$BEST values
represent the optimum magnitudes in the presence of neighboring
sources, they may still systematically underestimate the total flux of
extended sources by about 10\% \citep{mc05,cal08}. Therefore, we
corrected for this `lost' flux using the appropriate aperture
corrections. We used an approximate aperture radius
$r=(a^2+b^2)^{1/2}$ for our photometry by combining half the major
axis, $a$, and half the minor axis, $b$. We then calculated the
aperture corrections on the basis of template growth curves (derived
from the LGGS data) that were most representative of our extended
star clusters. The aperture corrections are slightly different for
different filters and different images: on average, they are
$\sim0.36$ mag for $r \le 3$ pix, $\sim0.10$ mag for $3<r\le5$ pix,
$\sim0.05$ mag for $5 < r \le7$ pix, and $\sim0.02$ mag for $r>7$
pix. The maximum aperture used for our photometry is 7.19 pix
($1''.85$). We therefore used $r = 7 \mbox{ pix} \approx1''.8$ as
the radius for our full sample's aperture corrections.

Many previous studies \citep[e.g.,][]{san09,san10,pl07} used a fixed
aperture of $r=2''.2$ for the photometry of all clusters and a smaller
aperture, $r \approx 1''.5$, for color measurements. For comparison,
based on the sizes and (elliptical) profiles of the clusters in our
sample, we used apertures of $r \le 1''.5$ for the photometry of 95\%
of our sample clusters. Nevertheless, the apertures adopted in both
this article and previous studies result in essentially the same
photometry and colors. The object names we use follow the naming
convention of \citet{san10} and SM10.

\begin{figure}
  \centerline{
    \includegraphics[scale=0.45,angle=0]{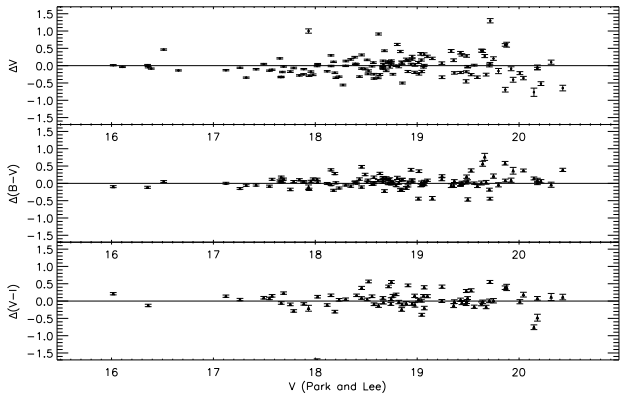}}
  \caption[]{Comparisons of our photometric measurements with those of
    \citet{pl07} for all star clusters in common. The error bars
    represent a combination (addition in quadrature) of the
    uncertainties associated with our photometry and those from the
    literature.}
  \label{fig2}
\end{figure}

Previously, \citet{pl07} determined integrated $BVI$ aperture
photometry for 104 M33 star clusters using (mainly) CFHT images, as
well as supplementary {\sl HST}/WFPC2 archival images. Charge-transfer
(in)efficiency (CTE) corrections were applied to the {\sl HST}
magnitudes, and all photometry has been converted to the standard
Johnson--Cousins system. Figure~\ref{fig2} compares our photometry
with that of \citet{pl07}, which shows that there is little systematic
difference: $\Delta V=0.006\pm0.001$ mag \citep[in the sense, this
  paper minus][]{pl07}, with $\sigma=0.349$ mag. Both our $(B-V)$ and
$(V-I)$ colors show good agreement with \citet{pl07} down to the
faintest magnitudes. The differences between both sets of photometry
are $\Delta (B-V)=0.049\pm0.004$ mag with $\sigma=0.189$ mag, and
$\Delta (V-I)=0.025\pm0.003$ mag, $\sigma=0.302$ mag.

\begin{figure}
\centerline{
  \includegraphics[scale=0.45,angle=0]{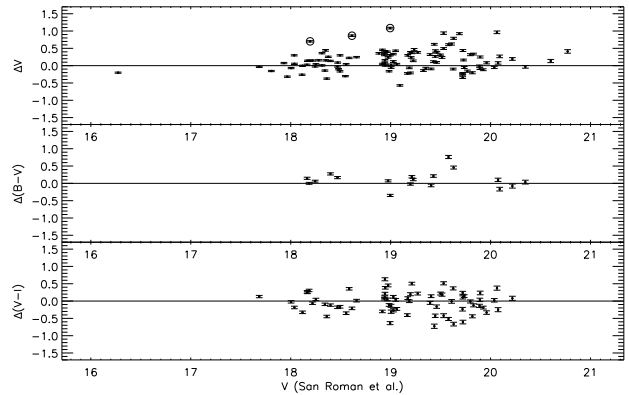}}
\caption[]{As Fig.~\ref{fig2} but for our photometry and that of
  \citet{san09}. Since \citet{san09} do not provide their photometric
  uncertainties, the error bars only reflect the uncertainties in our
  photometry. The open circles are objects 105, 110, and 148 of
  \citet{san09}.}
\label{fig3}
\end{figure}

Figure~\ref{fig3} shows the difference between our photometry and that
of \citet{san09}, whose database includes integrated photometry of 161
star clusters in M33 based on {\sl HST}/ACS--WFC observations. The
photometric uncertainties associated with the mean offsets are defined
as $\sigma/\sqrt{N}$, where $\sigma$ is the standard deviation and $N$
the number of data points. CTE corrections were applied and the
photometry has been converted to the standard Johnson--Cousins
system. Figure~\ref{fig3} shows comparisons of the respective sets of
magnitudes and colors. Since \citet{san09} do not provide their
photometric uncertainties, the error bars in Fig. \ref{fig3} reflect
our photometric uncertainties only. Our $V$-band magnitudes are in
good agreement with those of \citet{san09} down to the faintest
magnitudes: $\Delta V=0.162\pm0.016$ mag \citep[this paper
  minus][]{san09}, with $\sigma=0.304$ mag. However, a few objects
have magnitudes that are $>0.5$ mag fainter in our database than in
the tables of \citet{san09}, i.e., objects 105, 110, and 148 of
\citet{san09}, which are marked with open circles in
Fig. \ref{fig3}. We checked the $V$-band images and found that all of
these objects are located close to a few fainter neighboring sources,
which are treated as independent objects in our photometry but they
were regarded as stars belonging to the same cluster by
\citet{san09}. Both our $(B-V)$ and $(V-I)$ colors show good agreement
with \citet{san09}. The differences between our measurements and their
photometry are $\Delta (B-V)=0.106\pm0.025$ mag with $\sigma=0.233$
mag and $\Delta (V-I)=-0.065\pm0.008$ mag, $\sigma=0.361$ mag.

\begin{figure}
\centerline{
  \includegraphics[scale=0.45,angle=0]{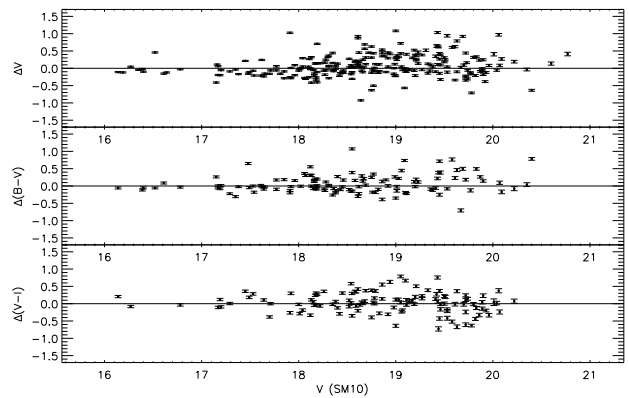}}
\caption[]{As Fig.~\ref{fig3} but for our photometry and that of
  SM10. Since SM10 do not provide their photometric uncertainties, the
  error bars only reflect the uncertainties associated with our
  photometry.}
  \label{fig4}
\end{figure}

We also compared our photometry with the measurements of SM10, who
assembled their photometry from the recent
literature. Figure~\ref{fig4} shows the relevant comparisons. Since
SM10 do not provide their photometric uncertainties, the error bars
only reflect the uncertainties associated with our photometry. Our
photometry is generally consistent with that of SM10: $\Delta
V=0.082\pm0.005$ mag (in the sense of this paper minus SM10) with
$\sigma=0.322$ mag. Again, both the $(B-V)$ and $(V-I)$ colors show
good agreement with SM10 down to the faintest magnitudes. The
differences between our colors and their photometry are $\Delta
(B-V)=0.069\pm0.006$ mag with $\sigma=0.267$ mag and $\Delta
(V-I)=0.012\pm0.001$ mag, $\sigma=0.343$ mag.

\begin{figure}
\centerline{
  \includegraphics[scale=0.42,angle=0]{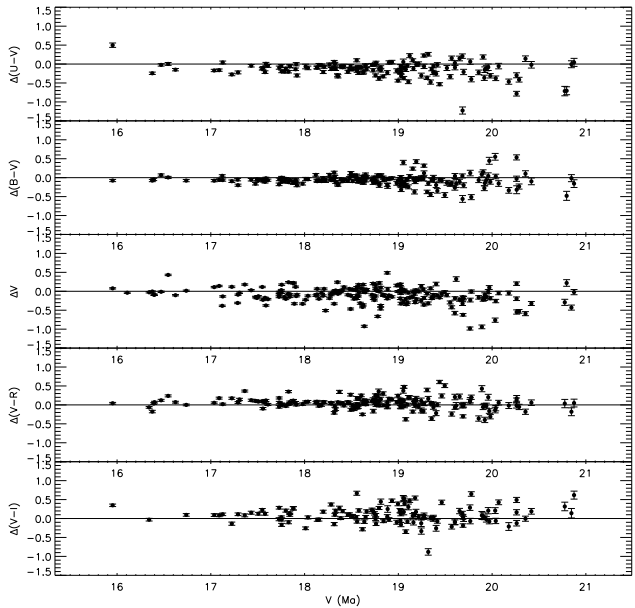}}
\caption[]{Comparison of our photometry and colors with the equivalent
  measurements of \citet{ma12}. The error bars are a combination
  (added in quadrature) of the uncertainties associated with our
  measurements and those of \citet{ma12}.}
  \label{fig5}
\end{figure}

We also compare our newly obtained photometric magnitudes and colors
with those published by \citet{ma12} in Fig.~\ref{fig5}. The error
bars shown in this figure are a combination (added in quadrature) of
the uncertainties associated with our measurements and those from the
literature. For all sources, the $V$-band offset is $\Delta
V=-0.100\pm0.007$ mag \citep[again, in the sense this paper
  minus][]{ma12}, with $\sigma=0.257$ mag. Our $V$-band magnitudes are
in good agreement with the equivalent values of \citet{ma12} for
bright sources, $V<18$ mag, while the photometry of \citet{ma12} seems
to be somewhat fainter than our photometry for $V>18$ mag. In fact,
\citet{ma12} noted that his $V$-band photometry is systematically
fainter than the previously published photometric measurements of
\citet{sm07}, \citet{pl07}, and \citet{san09}, so this result is in
line with our expectations. Our $(U-V)$, $(B-V)$, $(V-R)$, and $(V-I)$
colors show good agreement with the measurements of \citet{ma12}. The
differences between our and his colors are $\Delta
(U-V)=-0.151\pm0.013$, $\sigma=0.244$ mag; $\Delta
(B-V)=-0.068\pm0.005$, $\sigma=0.154$ mag; $\Delta
(V-R)=0.049\pm0.004$, $\sigma=0.157$ mag; and $\Delta
(V-I)=0.080\pm0.010$, $\sigma=0.274$ mag.

\begin{figure}
  \centerline{
    \includegraphics[scale=0.45,angle=0]{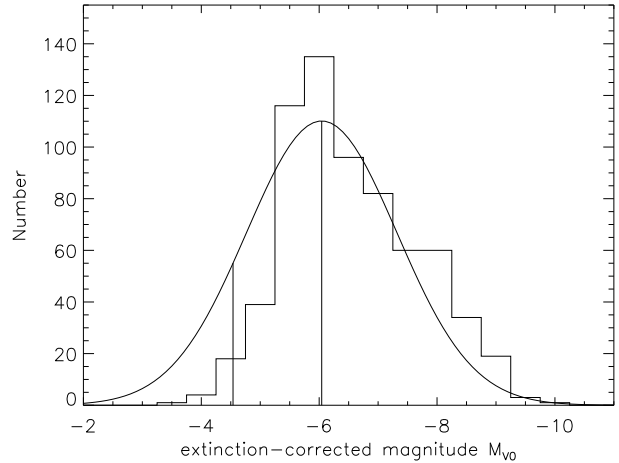}}
  \caption[]{Reddening-corrected absolute $V$-band magnitude ($M_V^0$)
    distribution of our M33 star cluster sample. The vertical line at
    $M_V^0= -4.54$ reflects our estimate of the sample's 50\%
    completeness limit.}
  \label{fig6}
\end{figure}

Figure \ref{fig6} shows the luminosity function of the M33 star
clusters and candidates in our sample, which can be used to estimate
the completeness of our photometry. The magnitudes are
extinction-corrected absolute $V$-band magnitudes. We adopted a
distance modulus to M33 of $(m-M)_0 = 24.64$ mag \citep{gal04}.
Extinction determinations for these star clusters were taken from
\citet{pl07} and \citet{san09}. For star clusters without published
reddening values, we adopted an average reddening of $E(V-I)=0.06$ mag
\citep{sa00,san09}, since for a significant number of clusters
deriving individual reddening values is not possible. In fact,
\citet{sa00} found that the standard deviation associated with the
average reddening value is $\Delta E(V-I)=0.02$ mag, which shows
that the scatter in reddening among most M33 star clusters is not
significant. Reddening variations may introduce a maximum additional
uncertainty of $\sim0.08$ mag in the $U$ band, and much smaller
uncertainties in the other, redder filters, particularly in the
near-infrared (NIR) bands. This is similar to the uncertainties in
our photometry. We thus conclude that variations in the average
reddening are unlikely to affect our fit results more significantly
for clusters without prior reddening estimates than the individual
reddening values determined previously for most of our other sample
clusters in M33.

Using a bin size of 0.5 mag, we determined an overall limiting
magnitude of $M_V^0= -4.54$ mag, which corresponds to the half-peak
height of the distribution; this follows the method adopted in our
previous analysis \citep{fan10} of M31 globular clusters (GCs). In
addition, we found that the peak of the distribution occurs at
$M_V^0=-6.04\pm0.04$ mag, with $\sigma_{M_V^0}=1.277$ mag, which we
adopt as the completeness magnitude threshold. Note that a Gaussian
`fit' does not seem to actually fit the data very well. It
underpredicts cluster numbers at the bright end and somewhat
overpredicts them at the faint end. Therefore, we give more weight to
the bright end in our fitting routine. The faint end may not be
Gaussian in (true) shape but simply be depleted because of sample
incompleteness.
  
\begin{figure*}
  \centerline{
    \includegraphics[scale=0.8,angle=0]{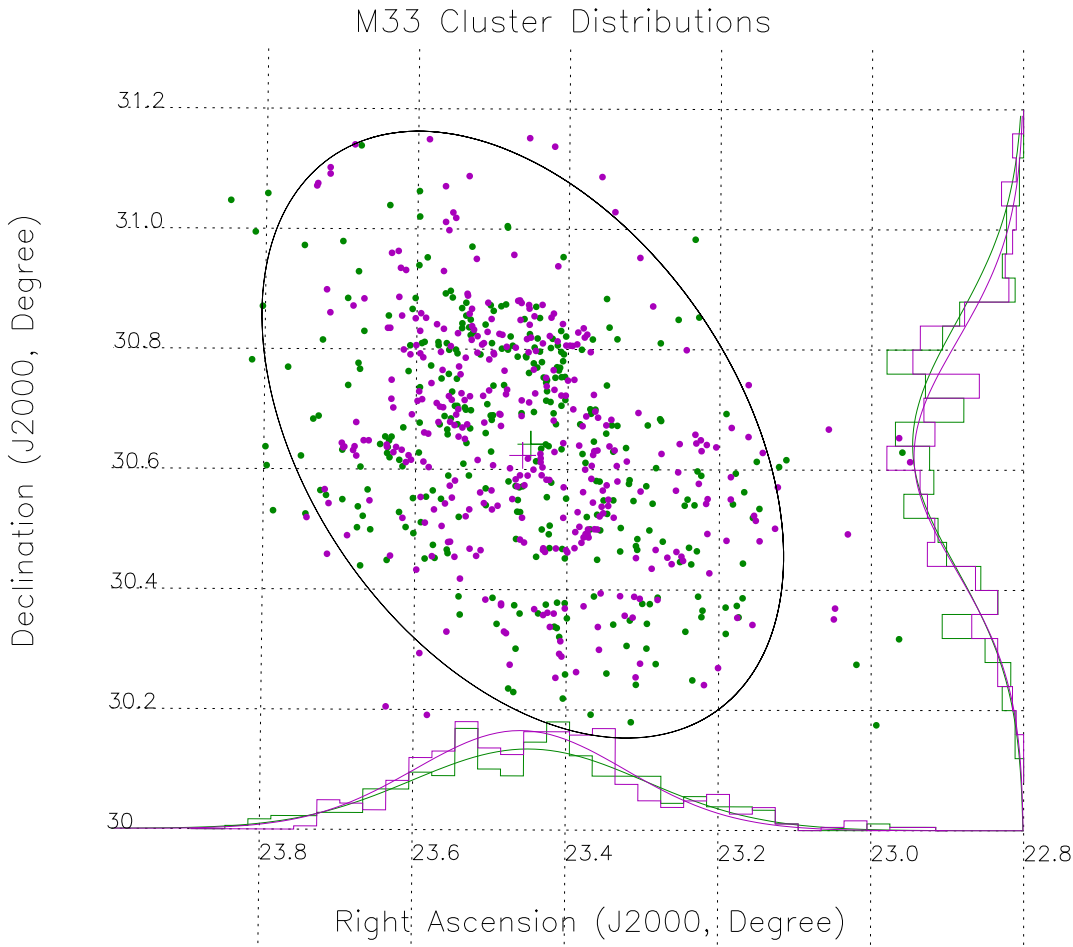}}
  \caption[]{Spatial distribution of our M33 star cluster sample. The
    green points represent clusters with $M_V^0 \le -6.04$ mag; the
    purple points are clusters with $M_V^0 >-6.04$ mag. The two plus
    signs represent the central positions of the Gaussian
    distributions for the two populations. The solid ellipse
    represents the galaxy's $D_{25}$ radius.}
  \label{fig7}
\end{figure*}

To further explore the completeness level of the M33 star cluster
sample, we show its spatial distribution in Fig.~\ref{fig7}. The
sample clusters characterized by absolute, extinction-corrected
$V$-band magnitudes, $M_V^0 \le -6.04$ mag, which is the peak magnitude
of the distribution in Fig. \ref{fig6}, are shown as green points,
while the purple points represent the $M_V^0>-6.04$ mag star clusters
in our sample. There is no evidence of any spatial differences between
both samples. We fit Gaussian profiles to the distributions in the
right ascension (R.A.) and declination (Dec) directions and determined
the distribution's center coordinates: R.A. = 23.449$^\circ$, with
$\sigma=0.153^\circ$, and Dec = 30.642$^\circ$, with
$\sigma=0.205^\circ$, for the bright sources; R.A. = 23.459$^\circ
(\sigma=0.138^\circ)$ and Dec = 30.623$^\circ (\sigma=0.192^\circ)$
for the faint objects. The two center positions are shown as the plus
signs in Fig. \ref{fig7}.

\section{SED Fits and Results}
\label{s:ana}

We constrain the ages, metallicities, and masses of the star clusters
based on SED fitting using $\chi^2$ minimization. We mainly used our
photometry (Table \ref{photo}) from the LGGS images to do so,
supplemented with the SDSS $ugriz$ photometry from
\citet{san10}. After elimination of those clusters for which
photometry is available in too few passbands,\footnote{We only apply
  SED fitting to clusters with photometric measurements in $\ge 3$
  passbands; measurements in fewer filters lead to highly unreliable
  results \citep[cf.][]{anders04b}.} our sample is reduced to 671 star
clusters and cluster candidates. We will use this subsample for SED
fitting and analysis in the following sections.

\subsection{Age estimates}
\label{s:age}

As is common in relation to most ground-based observations, we can
only access the {\it integrated} spectra and photometry of most
extragalactic star clusters. Therefore, the ages, metallicities, and
other physical parameters are obtained through analysis of the
integrated data. As a matter of fact, a strong age--metallicity
degeneracy would likely affect our analysis if only optical photometry
were available \citep{wor94,ar96,kaviraj07}. \citet{anders04b}
recommend to use NIR photometry if available. Inclusion of at least
one NIR passband can significantly improve the accuracy of the
resulting cluster parameters and partially break this degeneracy. In
addition, \citet{deg05} and \citet{wu05} showed that NIR colors can
greatly contribute to breaking the age--metallicity and
age--extinction degeneracies. Therefore, we will combine our $UBVRI$
photometry with $JHK$ photometry from 2MASS when available to
disentangle the degeneracies and obtain more accurate results.

We fit the SEDs with the evolutionary tracks derived from the updated
{\sc parsec} isochrones, assuming a \citet{chab} lognormal stellar
initial mass function
(IMF).\footnote{http://stev.oapd.inaf.it/cgi-bin/cmd} These {\sc
  parsec} isochrones are available for metallicities $0.0001 \le Z \le
0.06$ ($-2.2 \le {\rm [M/H]} \le +0.5$ dex) and for stellar masses in
the range $0.1 \le M/M_{\odot} \le 12$, with revised diffusion and
overshooting for low-mass stars and improvements in the interpolation
scheme. Note that, at present, the {\sc parsec} isochrones do not
include the thermally pulsing asymptotic giant-branch (TP-AGB)
phase. We adopted 24 metallicities from $Z= 0.0001$ to 0.06,
essentially equally spaced in in $\log Z$ space. The maximum age for a
reliable interpolation in metallicity is 13.5 Gyr, or $\log( t \mbox{
  yr}^{-1})=10.13$. Therefore, we adopted 71 equally spaced time steps
from $\log( t \mbox{ yr}^{-1})=6.6$ (4 Myr) to $\log( t \mbox{
  yr}^{-1})=10.1$ (12.6 Gyr) in steps of $\Delta \log( t \mbox{
  yr}^{-1}) =0.05$. The models return isochrone tables and integrated
SSP magnitudes for a number of photometric systems. We adopted the
SDSS $ugriz$, Johnson--Cousins $UBVRI$, and 2MASS $JHK$ systems.

The magnitudes were corrected for reddening (obtained previously; see
above) assuming a \citet{ccm} extinction curve. Since the wavelength
ranges covered by the SDSS $ugriz$ and Johnson--Cousins $UBVRI$
systems are essentially the same, it is not necessary to use both for
our SED fitting. We assigned priority to using the broad-band
Johnson--Cousins $UBVRI$ photometry, since these filters have wider
bandwidths and, hence, the photometry could potentially have higher
higher signal-to-noise ratios, all else being equal. Where broad-band
Johnson--Cousins magnitudes were not available, the SDSS $ugriz$
photometry was used. Thus, the cluster ages ($t$) could be determined
by comparing, in the $\chi^2$ sense, the {\sc parsec} SSP synthesis
models with the observed SEDs and adopting $Z$ as a free parameter,
i.e.
\begin{equation}
\chi^2_{\rm min}(t,Z)={\rm
  min}\left[\sum_{i=1}^8\left({\frac{m_{\lambda_i}^{\rm
        obs}-m_{\lambda_i}^{\rm mod}}{\sigma_i}}\right)^2\right],
\label{eq2}
\end{equation}
where $m_{\lambda_i}^{\rm mod}(t,Z)$ is the integrated magnitude in
the $i^{\rm th}$ filter of a theoretical SSP at age $t$ and for
metallicity $Z$, $m_{\lambda_i}^{\rm obs}$ represents the observed,
integrated magnitude in the same filter,
$m_{\lambda_i}=UBVRI,ugriz,JHK$ when 2MASS data is available or
$m_{\lambda_i}=UBVRI,ugriz$ when 2MASS data is not available (all
magnitudes were transformed to the AB magnitude system for our SED
fits). The errors, which are used as weights ($=1/\sigma^2$) by the
fitting routine, are calculated as
\begin{equation}
\sigma_i^{2}=\sigma_{{\rm obs},i}^{2}+\sigma_{{\rm mod},i}^{2}.
\label{eq3}
\end{equation}
Here, $\sigma_{{\rm obs},i}$ is the observational uncertainty;
$\sigma_{{\rm mod},i}$ represents the uncertainty associated with the
model itself, for the $i^{\rm th}$ filter. Following \citet{deg05},
\citet{wu05}, \citet{fan06}, \citet{ma07,ma09}, \citet{fan10}, and
\citet{wang10}, we adopt $\sigma_{{\rm mod},i}=0.05$ mag.

The estimated ages with $1\sigma$ errors of the M33 star clusters are
listed in Table~\ref{agemass}. We estimated the uncertainty associated
with a given parameter by fixing all other parameters to their best
values, then varied the parameter of interest, and recorded an error
corresponding to the $1\sigma~\chi^2_{\rm min}$ value.

\begin{figure*}
  \centerline{
    \includegraphics[scale=0.45,angle=0]{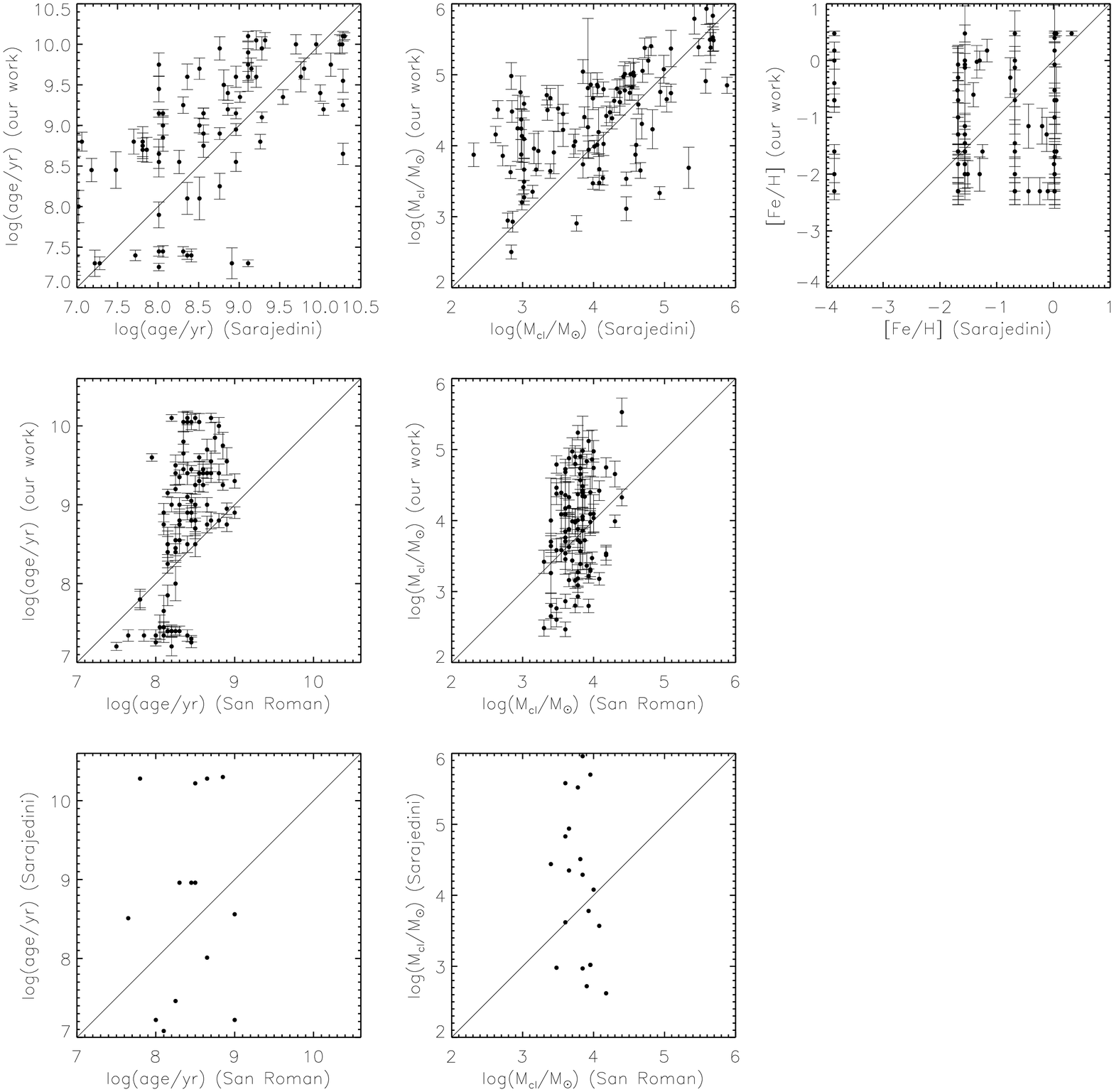}}
  \caption[]{Comparisons of our age, metallicity, and mass estimates
    with those of (top) SM10 and (middle) \citet{san09}. We also
    compare the parameter estimates of SM10 and \citet{san09} in the
    bottom row.}
  \label{fig8}
\end{figure*}

Our newly estimated ages, compared with previous results from the
literature, are plotted in Fig. \ref{fig8}. The top panels are
comparisons of SM10 and our results, and the middle panels are those
for our estimates and the results of \citet{san09}. The latter authors
compare MSTO photometry in the observed CMD with \citet{gir00}
theoretical isochrones assuming a metal abundance $Z=0.004$, which is
the mean of the disk abundance gradient in M33
\citep[cf.][]{kim,sa06,san09}. In the top panels, we note that SM10's
cluster ages and masses exhibit a general one-to-one trend with
respect to our results, although the scatter is relatively
large. In fact, all cluster ages and masses in SM10 were taken
from the Ma et al. series of articles. These latter authors derived
these parameters based on SED fits based on the Bruzual \& Charlot
SSP synthesis models (BC96). This is a similar approach to our
method, but based on different theoretical models and photometry. In
this article, we use some of the most up-to-date SSP synthesis
models currently available, while our photometry is based on (more
recent) observations with the Kitt Peak 4 m telescope. In addition,
we used NIR photometry here, which can be used successfully to
partially break the well-known age--metallicity degeneracy affecting
broad-band SED analyses. A comparison of the results of
\citet{san09} and the newly derived parameters presented here
reveals that the systematic differences, in a logarithmic sense, in
the ages, masses, and metallicities between our estimates and those
of SM10 (i.e., Ma et al.) are $-0.23\pm0.77$ dex, $-0.32\pm0.73$
dex, and $0.08\pm1.49$ dex, respectively. This scatter is partially
caused by the different metallicities adopted and partially because of
stochastic sampling effects \citep[see, e.g.,][]{anders13}. However,
the ages from \citet{san09} are systematically younger for clusters we
determine as $>1$ Gyr old, while they are systematically older than
our estimates for objects we return as $<0.1$ Gyr old. This may be
partially caused by the different analysis methods or by application
of different SSP models with different IMFs. 

To check if our method introduces any biases, we also compare the fit
results for \citet{san09} and those of SM10. This comparison shows the
same trends as seen in the comparisons of \citet{san09} and our
results, which indicates that the CMD fitting method of \citet{san09}
may be affected by a systematic bias (see the bottom panels of
Fig.~\ref{fig8}). Since \citet{san09} adopted a single metal
abundance for their CMD fits, we do not show a comparison of their
metallicities with those of SM10. In fact, as \citet{san09} point
out, the {\it isochrone-derived} ages for clusters younger than
$\sim1$ Gyr exhibit very little sensitivity to the assumed metal
abundance. This implies that any metallicity difference between our
results and those of \citet{san09} will likely have insignificant
effects on the final cluster parameters derived, since almost all
clusters in the sample of \citet{san09} are younger than 1 Gyr. Even
if \citet{san09} had allowed their metal abundances to vary, their
fit results would unlikey be affected significantly. Note that this
is rather different in comparison with SED fits. In addition, we
adopted the extinction values of \citet{san09}.

\citet{ppl09} compared cluster ages derived from resolved CMDs with
those from integrated photometry and found that ages derived from
resolved CMDs cover a relatively smaller range---$7.0 \la \log(t
\mbox{ yr}^{-1}) \la 8.5$---than those estimated based on integrated
colors and SEDs \citep[see,
  e.g.,][]{cha99,cha02,ma01,ma02a,ma02b,ma02c,ma04a,ma04b}. In
addition, \citet{sf97} found that stochastic sampling effects can
strongly affect the integrated $VJHK$ magnitudes of star clusters with
ages $7.5<\log(t \mbox{ yr}^{-1})<9.25$, in particular for less
massive clusters \citep{buz89,bj12,sl11}.

In fact, \citet{sl11} discussed the effects of stochastic sampling on
CMD fits; such effects are particularly prominent for low-mass star
clusters. Since star clusters are composed of finite numbers of stars,
stochastic sampling of the stellar IMF can significantly affect the
derived integrated cluster parameters, such as their ages,
metallicities, and masses. Recently, \citet{anders13} quantified the
effects of stochastic sampling of stellar IMFs based on a set of GALEV
SSP models for a wide range of (input) masses, metallicities,
foreground extinction values, and photometric uncertainties for their
model star clusters. They derive the accuracy of the integrated
parameter determination in different age ranges based on performing
fully sampled integrated-SED fits. For low-mass ($\sim10^3 M_{\odot}$)
clusters that are older than 10 Myr, the dispersion in $\log(t \mbox{
  yr}^{-1})$ could be as much as $\sim1$--2 dex, while the dispersion
in $\log(M_{\rm cl}/M_\odot)$ could be of the same order for [Fe/H] =
0.0 dex, foreground $E(B−V) = 0.0$ mag, and assuming photometric
uncertainties of 0.1 mag for all $UBVRIJHK$ magnitudes. In addition,
using a variety of metallicities and different combinations of
photometric passbands, stochastic sampling effects can even lead to
differences of $\Delta \log(t \mbox{ yr}^{-1}) >1$ dex. The offset
between the estimates of SM10 and \citet{san09} can thus be understood
easily. This type of effect could also partially explain the offsets
in the derived parameters between SM10 and our determinations, as well
as those between SM10 and \citet{san09}.

We emphasize that \citet{san09} also point out that, since their
photometry is generally not deep enough to detect the MSTO, few of
their clusters are returned as older than 1 Gyr. The middle panel in
the central row of Fig.~\ref{fig8} shows a comparison of the mass
estimates of \citet{san09} with our new determinations. We note that
the masses derived by \citet{san09} range from $5\times10^3 M_{\odot}$
to $5\times10^4 M_{\odot}$, while our estimates cover the range from
$\sim10^2 M_{\odot}$ to $\sim10^6 M_{\odot}$. This difference can
largely be traced back to the differences in our derived ages.

\begin{figure*}
  \centerline{ \includegraphics[scale=0.4,angle=0]{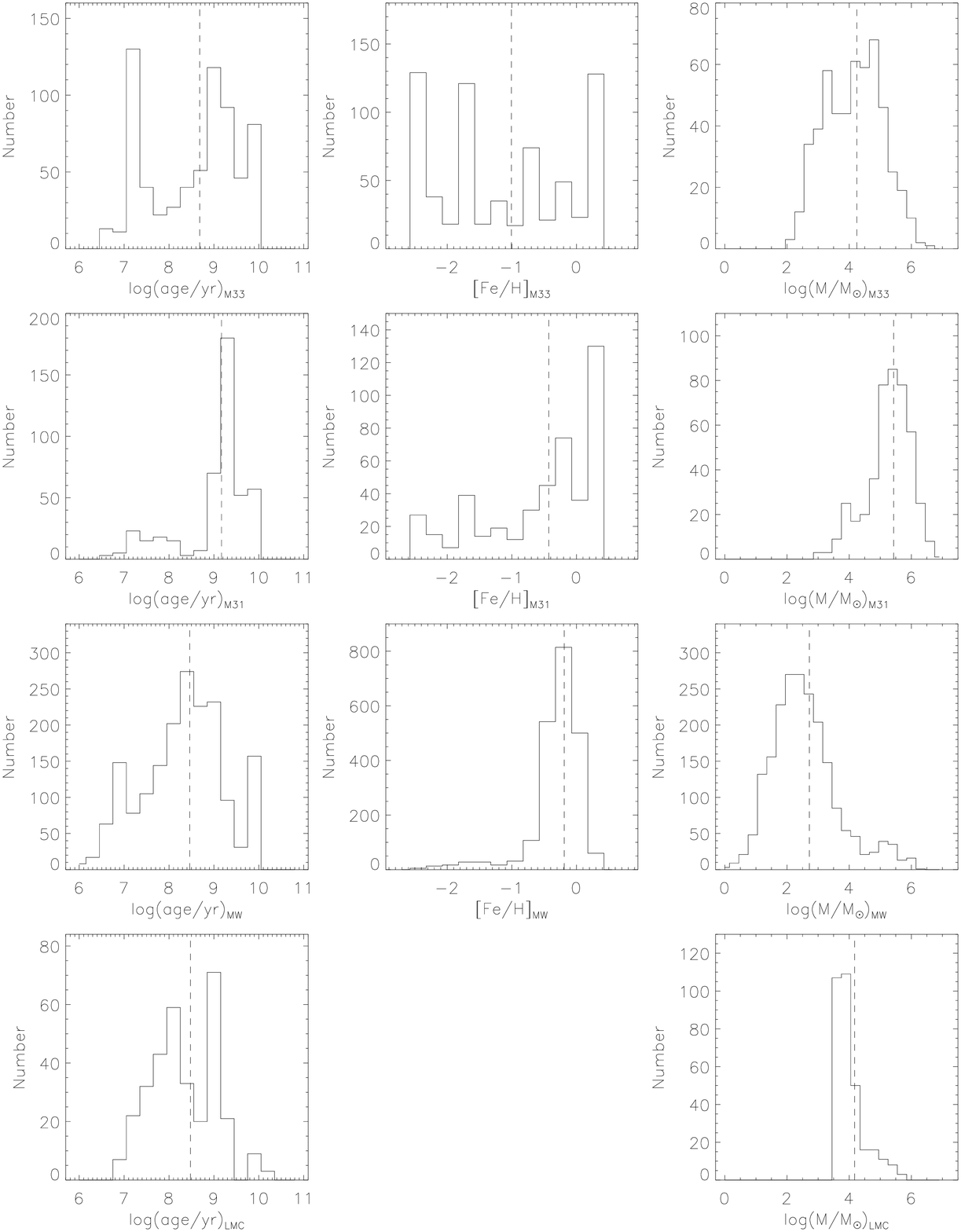}}
  \caption[]{Estimated age, mass, and metallicity distributions of the
    star clusters and candidates in M33, M31, and the Milky Way based
    on the {\sc parsec} models. The vertical dashed lines
    represent the mean values of the distributions. The photometric
    data of the M33 clusters comes from this paper, while the SEDs of
    the M31 globular-like clusters were analyzed by \citet{fan12}. The
    Milky Way star clusters are composed of GCs and OCs, for which the
    distributions are based on \citet{har10}, the updated catalog of
    \citet{dias02}, and \citet{go97}. The distributions of the LMC
    star clusters are based on \citet{bau13}.}
  \label{fig9}
\end{figure*}

The left-hand panels in the top and second rows of Fig.~\ref{fig9}
show the age distribution of our sample clusters in M33, as well as
the distribution of a representative sample of M31 star clusters
\citep{fan12}, for a bin size of 0.3 dex. To allow a reasonable
comparison, the parameters, such as age, metallicity, and mass of the
M31 star clusters were also all derived based on the {\sc parsec}
models, using the photometric data from \citet{fan12}. We also plot
the age distribution of the Milky Way star cluster sample, which is
composed of both GCs and open clusters (OCs). In addition, the age
estimates of LMC star clusters taken from \citet{bau13} are also
plotted in this diagram, for comparison. We note that the M33 star
clusters in our sample exhibit two peaks, at ages of $\sim10$ Myr and
$\sim1$ Gyr. The mean ages of the cluster samples are $\log(t \mbox{
  yr}^{-1})=8.68$, 9.17, 8.46, and 8.48 for M33, M31, MW, and the LMC,
respectively. For M33, the clusters with ages in excess of 10 Gyr were
most likely created during the epoch when the galaxy formed, while the
young star clusters might have been created in a number of mergers
during the last few Gyr or by the postulated recent galactic encounter
with M31 a few Gyr ago, suggested by \citet{mc09}. The age
distribution of the star clusters in M31 is dominated by clusters with
ages between 1 Gyr and the {\sl WMAP9} age of $13.77\pm0.06$ Gyr
\citep{ben12}. The age distribution of the Milky Way star clusters is
based on the combination of that of the OCs collected in
\citet{dias02}---the {\it New Catalog of Optically Visible Open
  Clusters and Candidates}; version 3.3---and the GCs, for which we
assumed the {\sl WMAP9} age. The mean age of the Milky Way's OCs is
similar to that of the M33 and LMC clusters. This latter similarity
implies that both galaxies have recently gone through one or more
periods of active star (cluster) formation. It is clear that there is
a much higher fraction of young star clusters in M33 (30.6\% are older
than 2 Gyr) than in M31, where 55.8\% of the clusters are older than 2
Gyr.

The middle panels of Fig.~\ref{fig9}, from the top to the third row,
show the metallicity distributions of star clusters in M33, M31, and
the Milky Way for a bin size of $\Delta \rm [Fe/H]=0.25$ dex. The mean
values of the three distributions are $\rm \overline{[Fe/H]}=-1.01$,
$-0.43$, and $-0.19$ dex. The metallicity distribution of the M33 star
clusters comes from our SSP fits based on the {\sc parsec} models,
while the metallicity distribution of the M31 star clusters is from
the data of \citet{fan12}, also based on the {\sc parsec} models. The
metallicity distribution of the Milky Way GCs has been plotted based
on the GC catalog of \citet{har10} and that of the OCs is based on the
\citep{dias02} catalog, which was most recently updated in 2013 and
includes 201 Milky Way OCs with metallicity measurements. We computed
the weighted metallicity of the two cluster samples using the numbers
of GCs and OCs as weights, which is therefore dominated by the
metallicity distribution of the OCs.

\subsection{Masses of the M33 star clusters}
\label{s:mass}

We calculated the clusters' theoretical mass-to-light ratios ($M/L_V$)
using the {\sc parsec} models, luminosities based on conversion of the
$V$-band fluxes, and a distance modulus of $(m-M)_0 = 24.64$ mag. The
resulting masses are listed in Table~\ref{agemass}. Figure \ref{fig9}
(right column) shows the mass distribution of the M33 star clusters in
our sample, as well as the masses of the M31 and Milky Way
clusters. The masses of the Galactic GCs were calculated by
\citet{go97}, while those of 650 OCs with mass estimates are from
\citet{pis08}. As before, these Galactic cluster mass estimates were
combined, using as weights the total numbers of clusters of different
types. For comparison, we also include the LMC star cluster data from
\citet{bau13}. The bin size is $\Delta \log(M_{\rm cl}/M_\odot)=0.3$
dex. The mean mass of the M33 clusters is $\log(M_{\rm
  cl}/M_\odot)=4.25$ ($1.78\times10^4M_{\odot}$) while the mean values
for the M31 star clusters, the combined sample of Galactic star
clusters, and the LMC clusters are $\log(M_{\rm cl}/M_\odot)=5.43$
($2.69\times10^5M_{\odot}$), $\log(M_{\rm cl}/M_\odot)=2.72$
($5.24\times10^2M_{\odot}$), and $\log(M_{\rm cl}/M_\odot)=4.18$
($1.51\times10^4M_{\odot}$), respectively. Figure \ref{fig9} shows
that the mean mass of the star clusters in M33, which is similar to
that of the LMC clusters, is much lower than the equivalent masses in
M31 and the Milky Way, suggesting that the M33 cluster population on
the whole is dominated by lower-mass clusters.

\begin{figure}
\centerline{
  \includegraphics[scale=0.42,angle=0]{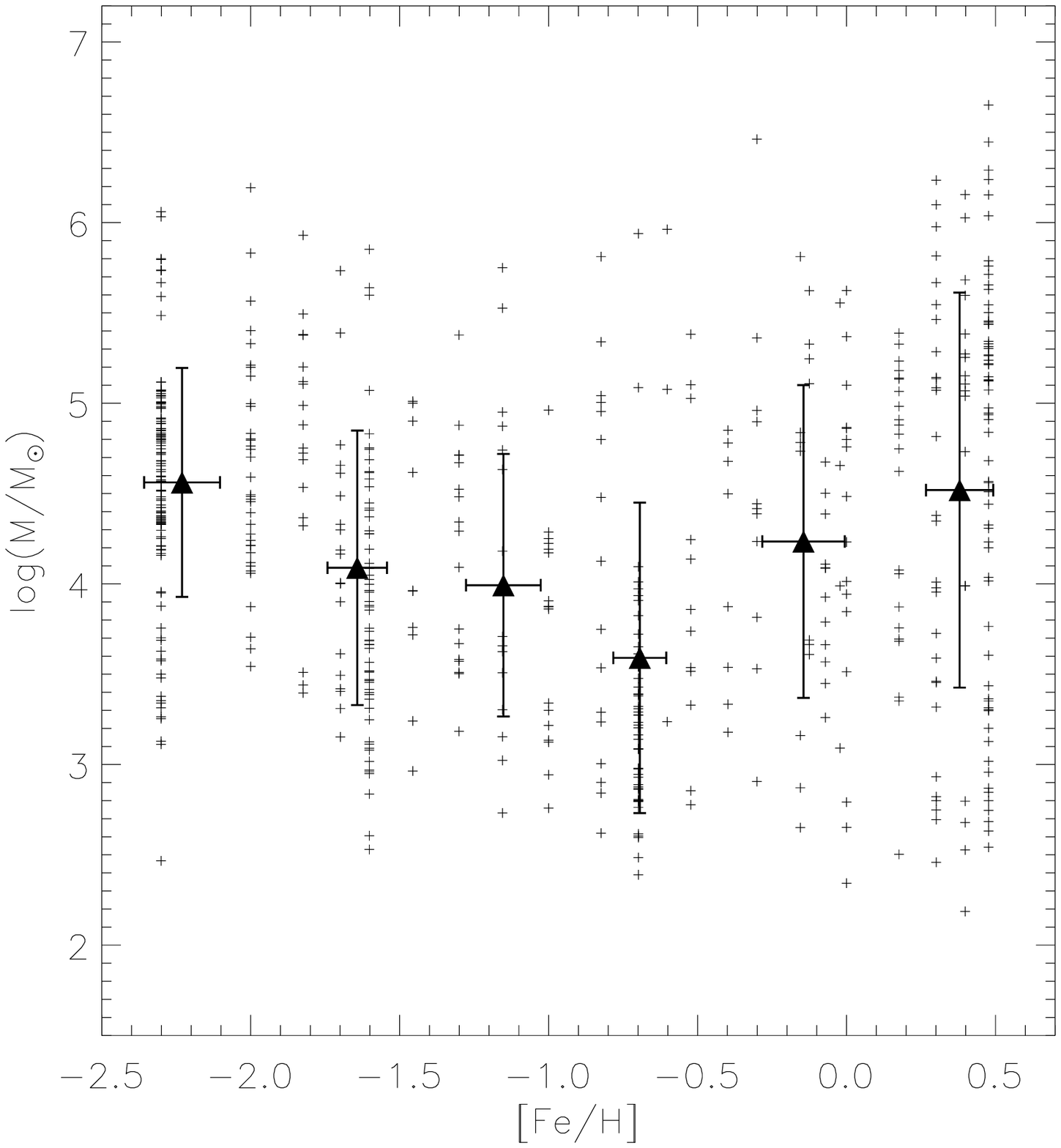}}
\caption[]{Masses versus metallicities for the M33 star clusters. We
  divided the metallicities into 6 bins, each with a size of 0.5
  dex. The filled triangles with error bars represent the mean values
  and $\sigma$'s for each bin.}
  \label{fig10}
\end{figure}

The mass--metallicity relation (MMR) for star clusters in the `blue
sequence' (which is known as the `blue tilt') has been discussed for
many external galaxies, e.g., for a sample of six giant elliptical
galaxies \citep{ha09a}, M87 \citep{peng09,ha09b}, the Sombrero galaxy
\citep{ha09c}, and M31 \citep{fan09}. Self-enrichment was considered a
reasonable explanation by both \citet{bh09} and \citet{ss08}, who
suggested that the level of star formation is controlled by supernova
feedback, and the efficiency scaling is proportional to the
proto-cloud mass. Since we have derived the masses and metallicities
of the M33 clusters, we can investigate their MMR. In
Fig. \ref{fig10}, we plot cluster masses as a function of metallicity
for our sample star clusters. The filled triangles with associated
error bars represent the mean values and $\sigma$'s for each bin. The
bin size is 0.5 dex in metallicity. For low metallicities, ${\rm
  [Fe/H]}<-0.8$ dex, the cluster masses seem to decrease with
metallicity, while for ${\rm [Fe/H]}\ge-0.8$ dex, the cluster masses
increase with increasing metallicity.

\begin{figure}
  \centerline{
  \includegraphics[scale=0.42,angle=0]{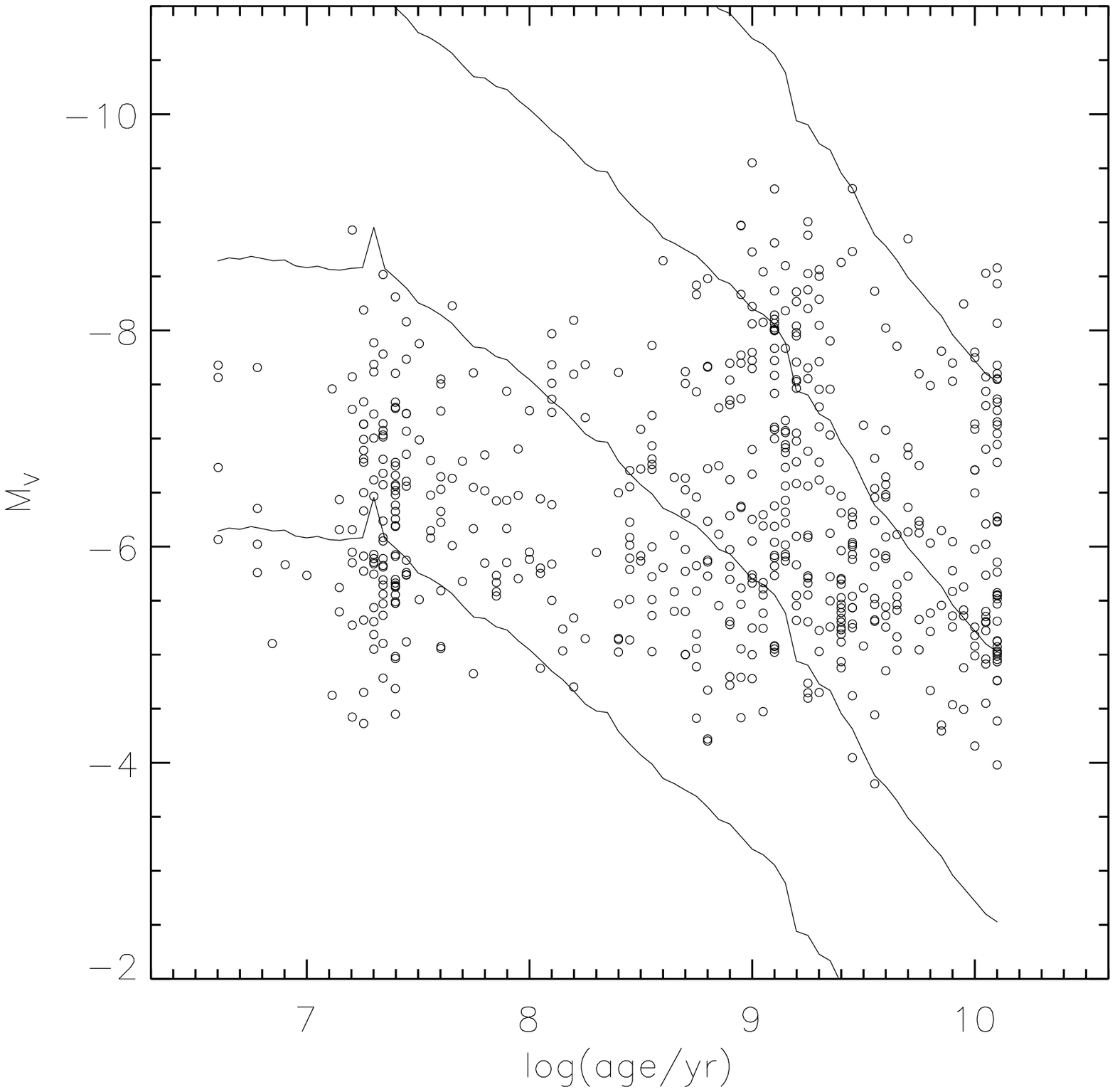}}
\caption[]{Extinction-corrected absolute magnitude as a function of
  cluster age. The solid lines are the expected relations for SSPs
  from the {\sc parsec} model with $Z = 0.004$ and masses of $10^3,
  10^4, 10^5$, and $10^6 M_{\odot}$ (from bottom to top).}
\label{fig11}
\end{figure}

Figure~\ref{fig11} shows the extinction-corrected absolute magnitudes
as a function of age for our sample star clusters. The solid lines
represent theoretical isochrones from the updated {\sc parsec} models
for masses of $10^3, 10^4, 10^5$, and $10^6M_{\odot}$, for a
metallicity of $Z =0.004$. The masses of most star clusters and
candidates are between $10^3 M_{\odot}$ and $10^5 M_{\odot}$. The less
massive clusters, $M_{\rm cl}<10^3 M_{\odot}$, tend to be young ($<2$
Gyr) while clusters with $M_{\rm cl} > 10^5 M_{\odot}$ are generally
old ($>2$ Gyr).

\begin{figure}
\centerline{
  \includegraphics[scale=0.42,angle=0]{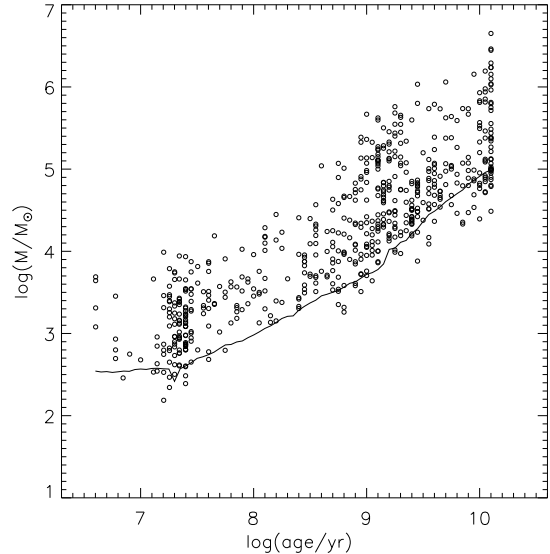}}
\caption[]{Mass as a function of age for the M33 star clusters. We
  also show the fading (`completeness') limit for $M_V \approx-5$ mag
  from Fig. \ref{fig6}, based on the {\sc parsec} SSP models for a
  metallicity $Z=0.004$.}
  \label{fig12}
\end{figure}

The age--metallicity relations of extragalactic star clusters have
been studied by many authors \citep[e.g.,][]{da06,wang10,fan10}.
Figure \ref{fig12} shows the age-versus-mass estimates of our M33
sample clusters, as well as the so-called `fading line,' which is
roughly equivalent to the $\sim 50$\% completeness limit. The
theoretical line is based on the {\sc parsec} SSP models for a
metallicity of $Z=0.004$. For the detection limit, an absolute
magnitude of $M_V\approx-5$ mag is estimated from
Fig.~\ref{fig6}. Indeed, most star clusters lie above the line. The
small number of clusters found below the fading line may be due to
either a possibly variable photometric completeness level or
underestimated extinction. A few faint, young ($<20$ Myr-old) clusters
could be young OCs. Note that there are three overdensity regions in
this figure, at (i) $\log(t \mbox{ yr}^{-1})\approx7.3$ (20 Myr), (ii)
$\log(t \mbox{ yr}^{-1})\approx9$ (1 Gyr), and (iii) the {\sl WMAP9}
age ($\sim13.7$ Gyr); the latter represents the subset of the cluster
population that seems to have formed during the time of the galaxy's
formation.

\section{Summary}
\label{s:sum}

We have obtained $UBVRI$ photometry for 708 star clusters and cluster
candidates in M33, which were selected from \citet{san10}, including
387, 563, 616, 580, and 478 objects in the $UBVRI$ bands,
respectively, of which 276, 405, 430, 457, and 363 did not have
previously published $UBVRI$ photometry. The {\sc SExtractor} code was
applied to derive the photometry from LGGS archival images, which
cover 0.8 deg$^2$ along the major axis of M33.

We compared our photometry with previous measurements, which showed
that our photometry is generally consistent with previous measurements
in all filters. The ages, metallicities, and masses of our sample
clusters were derived by comparison of their observed SEDs with {\sc
  parsec} SSP synthesis models. The fits show that only 205 of the 671
clusters (30.55\%) are older than 2 Gyr, which is a much smaller
fraction than that derived for the M31 globular-like clusters
(55.80\%), suggesting that the M33 cluster population is dominated by
young star clusters ($<1$ Gyr). We also note that the mean mass of the
M33 star clusters is $1.78\times10^4M_{\odot}$, which is much less
massive than that of the M31 sample ($2.69\times10^5M_{\odot}$) and
similar to that of LMC cluster population ($1.51\times10^4M_{\odot}$),
but higher than that of Milky Way star clusters (including both GCs
and OCs: $5.24\times10^2M_{\odot}$). This may be related to the fact
that the mass of M33 is lower that that of M31, and the gravitational
potential is not large enough to produce as many GCs as in
M31. Instead, the star-formation history of M33 may be similar to that
of the LMC. As for the Milky Way, the recent few Gyr have seen the
galaxy undergo quiescent evolution (a low star-formation rate).

On the other hand, the mean metallicity of the M33 clusters ($\rm
[Fe/H]=-1.01$ dex) is much lower than that of the M31 star clusters
($\rm [Fe/H]=-0.43$ dex) and also of the Milky Way star clusters ($\rm
[Fe/H]=-0.19$ dex), suggesting that its star-formation history has
been rather different from those of either M31 or the Milky Way. Based
on the cluster mass distributions we also found that the mean mass of
star clusters in M33 is similar to that in the LMC but much lower than
that in M31 and higher than that in the Milky Way, indicating that M31
underwent more violent star formation than either M33 or the LMC.
We also note that stochastic sampling effects can significantly
affect our SED fit results for low-mass clusters (i.e., $M_{\rm cl}
\le 10^3 M_{\odot}$), potentially leading to large differences in
integrated cluster ages, metallicities, and masses
\citep{anders13}. The effects of stochasticity in the clusters'
stellar mass functions become weaker with increasing cluster
mass. However, even for the highest-mass clusters, $M_{\rm cl}
\simeq 2 \times 10^5 M_{\odot}$, the uncertainties in the derived
logarithmic ages are 0.05--0.25 dex; the equivalent uncertainties
pertaining to the derived masses are 0.09--0.17 dex. Although these
uncertainties are sometimes significant, obtaining accurate,
high-resolution spectroscopic observations for statistically large
samples of extragalactic star clusters is often prohibitive,
particularly if observing time on significantly oversubscribed
(large) telescopes is sought. Broad-band imaging and parameter
determination based on sophisticated SED fits is the realistic yet a
poor man's alternative approach.

The broad-band SED uncertainties can be further reduced for specific
age ranges, e.g., for young clusters that exhibit H$\alpha$ emission,
which places additional constraints on the most likely age range. At
present, a number of teams are working on quantifying the effects of
stochastic sampling; although the underlying message is that such
effects may have significant implications in terms of the precision of
the derived physical parameters, we argue that a proper understanding
of one's uncertainties is of the utmost importance. One should keep
these issues in mind when dealing with physical cluster parameters
based on broad-band observations.

\acknowledgments

This publication makes use of data products from the Two Micron All
Sky Survey, which is a joint project of the University of
Massachusetts and the Infrared Processing and Analysis
Center/California Institute of Technology, funded by the National
Aeronautics and Space Administration and the U.S. National Science
Foundation. This research is supported by the National Natural Science
Foundation of China (NFSC) through grants 11003021, 11043006,
11073001, 11373003, and 11373010. ZF also acknowledges a Young
Researcher Grant from the National Astronomical Observatories, Chinese
Academy of Sciences, while RdG acknowledges research support from the
Royal Netherlands Academy of Arts and Sciences (KNAW) under its
Visiting Professors Programme.

%\clearpage

\LongTables

\clearpage
\pagestyle{empty}
%\setcounter{page}{0}
%\vspace*{1.5in}
\begin{deluxetable}{cccccccc}
%\rotate
%\tablenum{1}
\tablecolumns{8} \tablewidth{0pc} \tablecaption{New $UBVRI$ photometry
  of M33 clusters.\label{photo}}

\tablehead{\colhead{Our ID} & \colhead{SR ID} & \colhead{SM ID} &
  \colhead{$U$} &\colhead{$B$} &\colhead{$V$} &\colhead{$R$}
  &\colhead{$I$} \\ \colhead{} &\colhead{} &\colhead{} &\colhead{mag}
  &\colhead{mag} &\colhead{mag} &\colhead{mag} &\colhead{mag} }
\startdata 
    1&  163&     & $  19.353\pm   0.033$ & $  18.127\pm   0.020$ &        \nodata        &        \nodata        &        \nodata        \\
    2&  197&     & $  19.898\pm   0.035$ & $  18.751\pm   0.021$ &        \nodata        & $  17.691\pm   0.019$ & $  17.606\pm   0.029$ \\
    3&  203&     & $  19.951\pm   0.034$ & $  20.596\pm   0.028$ & $  19.954\pm   0.027$ & $  19.634\pm   0.027$ & $  19.204\pm   0.035$ \\
    4&  204&     & $  19.448\pm   0.033$ & $  19.523\pm   0.021$ & $  18.856\pm   0.020$ & $  18.384\pm   0.018$ & $  17.797\pm   0.029$ \\
    5&  284&     & $  19.945\pm   0.034$ & $  18.686\pm   0.020$ & $  17.539\pm   0.018$ & $  16.742\pm   0.017$ &        \nodata        \\
    6&  316&     & $  19.890\pm   0.034$ & $  18.702\pm   0.020$ & $  17.367\pm   0.018$ & $  16.427\pm   0.017$ &        \nodata        \\
    7&  346&     & $  20.331\pm   0.038$ & $  20.078\pm   0.025$ & $  19.400\pm   0.023$ & $  18.935\pm   0.021$ & $  18.310\pm   0.030$ \\
    8&  394&     & $  17.709\pm   0.032$ & $  17.406\pm   0.019$ & $  16.998\pm   0.018$ & $  16.214\pm   0.017$ & $  15.852\pm   0.028$ \\
\enddata
\tablenotetext{a}{`SR' and `SM' refer to the IDs of \citet{san10} and
  SM10, respectively.}
\end{deluxetable}

\clearpage
\pagestyle{empty}
%\setcounter{page}{0}
%\vspace*{1.5in}
\begin{deluxetable}{cccc}
%\rotate
%\tablenum{2}
  \tablecolumns{4} \tablewidth{0pc} \tablecaption{Ages, metallicities
    and masses of the M33 star clusters.
\label{agemass}}
\tablehead{\colhead{Object} &\colhead{Age}
  &\colhead{[Fe/H]} & \colhead{${\rm log(M_{cl})}$}\\
  \colhead{} &\colhead{(Gyr)} &\colhead{(dex)} & \colhead{[${\rm M_{\odot}}$]}} \startdata 
SR1001& $   2.510\pm   0.775$ & $  -2.301\pm   0.150$ & $   4.184\pm   0.144$ \\
SR1009& $   1.580\pm   0.260$ & $   0.477\pm   0.088$ & $   5.154\pm   0.175$ \\
SR1012& $   1.120\pm   0.259$ & $  -1.602\pm   0.122$ & $   4.148\pm   0.129$ \\
SR1013& $   0.006\pm   0.002$ & $   0.301\pm   1.049$ & $   2.688\pm   0.112$ \\
SR1020& $   1.580\pm   0.185$ & $  -0.824\pm   0.150$ & $   4.025\pm   0.100$ \\
SR1021& $   0.022\pm   0.008$ & $  -1.602\pm   0.122$ & $   3.032\pm   0.081$ \\
SR1022& $   3.160\pm   1.810$ & $  -2.301\pm   0.238$ & $   4.683\pm   0.244$ \\
SR1024& $   3.980\pm   1.400$ & $  -2.301\pm   0.150$ & $   4.565\pm   0.004$ \\
\enddata
\end{deluxetable}

\end{document}